\documentclass[journal]{IEEEtran}

\usepackage{cite}

\usepackage{graphicx}

\usepackage{amsmath}
\usepackage{amsfonts}
\usepackage{amssymb}

\usepackage{url}

\usepackage[font=footnotesize]{caption} 
\usepackage{subcaption} 
\usepackage{siunitx}
\DeclareSIUnit\dBm{dBm}
\usepackage{collcell}

\usepackage{booktabs}       

\usepackage{listings}       
\usepackage{xcolor}

\definecolor{codegreen}{rgb}{0,0.6,0}
\definecolor{codegray}{rgb}{0.5,0.5,0.5}
\definecolor{codepurple}{rgb}{0.58,0,0.82}
\definecolor{backcolour}{rgb}{0.95,0.95,0.92}

\lstdefinestyle{mystyle}{
    backgroundcolor=\color{backcolour},   
    commentstyle=\scriptsize\color{codegreen},
    keywordstyle=\color{magenta},
    stringstyle=\color{codepurple},
    basicstyle=\ttfamily\footnotesize,
    breakatwhitespace=false,         
    breaklines=true,                 
    captionpos=b,                    
    keepspaces=true,                 
    showspaces=false,                
    showstringspaces=false,
    showtabs=false,                  
    tabsize=2
}
\usepackage{hyperref}		        
\usepackage[noabbrev]{cleveref}		

\begin{document}
\title{PRECISE\\Photonic hybRid EleCtromagnetIc SolvEr}
\author{Davide~Bazzanella,
        Mattia~Mancinelli,
        Massimo~Borghi,
        Paolo~Bettotti,
        and Lorenzo~Pavesi
\thanks{
This project has received funding from the European Research Council (ERC) under the European Union’s Horizon 2020 research and innovation programme (grant agreement No 788793, BACKUP),  and from the MIUR under the project PRIN PELM (20177 PSCKT)}
\thanks{D. Bazzanella, M. Mancinelli, P. Bettotti, and Lorenzo~Pavesi are with
the Department of Physics, University of Trento, Trento, TN 38123 Italy.}
\thanks{M. Borghi was with the Department of Physics, University of Trento, Trento, TN 38123 Italy.
He is now with the Department of Physics, University of Pavia, Pavia, 27100 Italy.}
\thanks{D. Bazzanella is the corresponding author and can be contacted at davide.bazzanella@unitn.it.}
}

%


\maketitle

\begin{abstract}
PRECISE, Photonic hybRid EleCtromagnetIc SolvEr, is a Matlab-based library to model large and complex photonic integrated circuits.
Each circuit is modularly described in terms of waveguide segments connected through multi-port nodes.
Linear, nonlinear, and dynamical phenomena are simulated by solving the system of differential equations describing the effect to be considered.
By exploiting the steady-state approximation of the electromagnetic field within each node device, the library can handle large and complex circuits even on desktop PCs.
We show that the steady-state assumption is fulfilled in a broad number of applications and we compare its accuracy with an analytical model (coupled mode theory) and experimental results.
PRECISE is highly modular and easily extensible to handle equations different from those already implemented and is, thus, a flexible tool to model the increasingly complex photonic circuits.
\end{abstract}


\IEEEpeerreviewmaketitle

\section{Introduction}
\IEEEPARstart{T}{he} continuous evolution of silicon photonic produces Photonic Integrated Circuits (PIC) of endlessly increasing complexity.
To model the optical response of large PICs requires intensive computational effort if approached by directly solving Maxwell equations with either FDTD or FEM methods.
Several numerical tools are already available to accurately model basic building blocks and simple PICs (e.g., MEEP \cite{oskooi2010meep}, FEM as Lumerical \cite{lumerical} and COMSOL \cite{comsol}, etc.).
Other open-source tools like PhotonTorch \cite{laporteHighlyParallelSimulation2019b}
and Symphony \cite{ploegSimphonyOpenSourcePhotonic2021} permit the analysis of linear systems only but lack the possibility to handle Non-Linear (NL) effects.
To our best knowledge, only CAPHE \cite{caphe} is able to handle nonlinear photonics devices.
Yet the analysis of complex circuits would greatly benefit from proper coarse-grain modeling, considering only the most representative aspects of PIC nodes and simplifying the circuit description to reduce the overall computational effort.
A significant simplification of the PIC descriptions will permit to increase the size of the circuit to be modeled as well as to include nonlinearities that are often at the heart of many photonic functions and nonlinearities are even increasing their interest in the emerging field of neuromorphic photonics \cite{sandeAdvancesPhotonicReservoir2017, caspaniIntegratedSourcesPhoton2017, makarovLightInducedTuningReconfiguration2017, borghiNonlinearSiliconPhotonics2017, xuSurveyApproachesImplementing2021}.
The description of NL optical phenomena requires solving systems of differential equations (ODEs) that are inherently serial problems and cannot efficiently exploit parallelization techniques (albeit recent libraries propose improvements \cite{rackauckas2017differentialequations}).

Here we propose the Photonic hybRid EleCtromagnetIc SolvEr (PRECISE) Toolkit to ease the modeling of linear and nonlinear spectra and the dynamics of complex PICs by assuming steady-state conditions of the electromagnetic fields within each fundamental circuit block.
We will show that the steady-state approximation is fulfilled in a large number of devices designs and for state-of-the-art applications and we confirm the excellent accuracy of our model by comparing its results with different analytical schemes and experimental results.
The modular structure of PRECISE does not rely on a specific assumption about the PIC building blocks and it handles any type of passive devices found in integrated photonics.
This fact enables the study and design of both known and novel photonic structures and to sweep over broad space parameters.

\section{Theoretical background and PICs description}
PRECISE describes PICs as waveguides (WGs) connected by nodes and interacting with the surrounding via I/O ports.
The idea of our code is to evaluate complex PICs without resorting to the direct solution of ODE systems at the sub \unit{\ps} time-scale.
For purely linear PICs, the system is described in terms of symbolic equations that are inverted and solved to calculate the parameters of interest.
In the case of nonlinear PICs, PRECISE assumes that the field intensity instantaneously achieves its steady-state, within each node.
Such approximation (similar to \cite{van2012simplified}) is satisfied whenever the optical field stationary state is reached on timescales smaller than the characteristic dynamics of the physical processes defining the nonlinear behavior (for example, the free carriers or the material’s temperature).
As we will show, this approximation is often fulfilled and greatly simplifies the solution of ODE systems.

\subsection{Description of PIC}
\begin{figure*}[htp]
    \centering
	\includegraphics{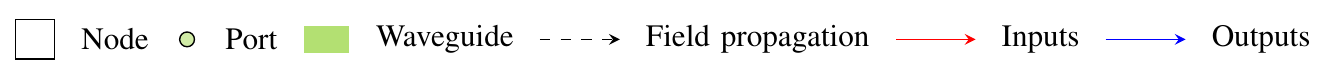}
	
	\vspace{.2em}
	\includegraphics{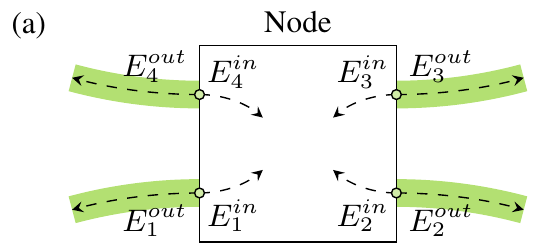}
	\includegraphics{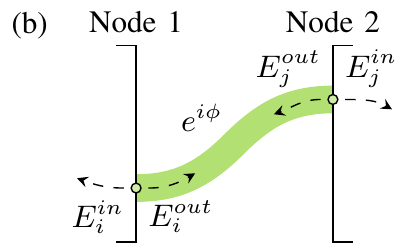}
	\includegraphics{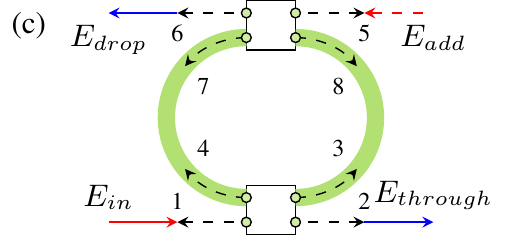}
    {\phantomsubcaption\label{fig:coupling_scheme}}
    {\phantomsubcaption\label{fig:waveguide}}
    {\phantomsubcaption\label{fig:ring-scheme}}
	
	\vspace{1em}
	\includegraphics{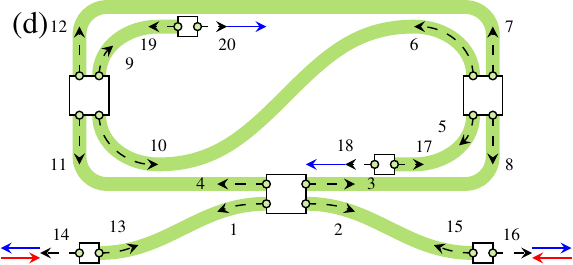}
	\includegraphics{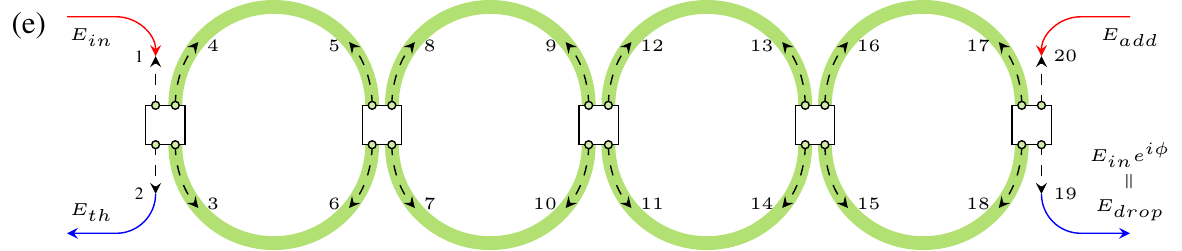}
    {\phantomsubcaption\label{fig:taiji-scheme}}
    {\phantomsubcaption\label{fig:crow-scheme}}
	\caption{
	Each system is firstly defined by a set of nodes (small squares), which group a number of ports (black dots).
	The WGs may then connect the ports and are represented by thick gray lines.
	The dashed arrows illustrate the input/output fields at each port and are identified by integer numbers.
	Input and output of the whole system are instead depicted as red and blue arrows respectively.
	\subref{fig:coupling_scheme} Example of a node with four ports, showing both input and output fields.
	\subref{fig:waveguide} Example of WG connecting two ports belonging to different nodes.
	\subref{fig:ring-scheme} Scheme of a MR resonator built with two $2\times2$ nodes, for a total of eight ports, and two WGs.
	    There are two input fields, \textit{input} and \textit{add}, and two output fields, \textit{through} and \textit{drop}.
	\subref{fig:taiji-scheme} Scheme of a Taiji MR.
	    This structure is composed of three $2\times2$ nodes and four $1\times1$ nodes, for a total of 20 ports, and eight WGs.
	\subref{fig:crow-scheme} Scheme of a CROW structure with four rings.
	    The structure is composed of five $2\times2$ nodes, for a total of 20 ports, and eight WGs.
	    The input/output fields are the same as the MR, but, in this case, both the \textit{input} and \textit{add} fields are used simultaneously.
	}
\end{figure*}

PICs are described programmatically by decomposing them into basic components: any PIC is essentially formed of WGs segments that are connected together at nodes.
To completely describe a PIC we need to define three fundamentals elements:
\begin{enumerate}
\item Nodes are logical blocks that groups a number of ports together.
    A node can have as many input/output ports as necessary and its generic connectivity is described by the following matrix equation:
    \begin{equation}
    	E^{out} = C \cdot E^{in}
    \end{equation}
    where $E^{in}$ and $E^{out}$ are $1\times n$ column vectors and $C$ is an $n \times n$ matrix.
    In its simplest form coupling only includes constant coupling parameters, but a frequency dependent coupling can also be included, thanks to the symbolic definition of the nodes themselves.
    For example, a $2\times2$ directional coupler is represented by a node with four ports (shown in \Cref{fig:coupling_scheme}).
    Its coupling matrix is
    \begin{equation}
        C =
        \begin{bmatrix}
            0 & t & k & 0 \\
            t & 0 & 0 & k \\
            t & 0 & 0 & k \\
            0 & k & t & 0 \\
        \end{bmatrix}.
    \end{equation}
    The user can add a node \texttt{add\_node} and define the coupling coefficients of its ports using \texttt{add\_link}, as it will be explained better later.
    
\item WG segments connect the output port (i) of a node to an input port (j) of the same or different node, as described in \Cref{fig:waveguide}.\\
We consider the solutions of the wave equation as a monochromatic plane wave propagating along the WG ($z$):
\begin{equation}
    U(x,y,z) = A \, u(x,y)\exp(-i\beta z) \,,
    \label{eq:plane_wave}
\end{equation}
whose transverse amplitude profile u($x$, $y$) and propagation constant $\beta$ belong to one of the guided modes.
In principle a WG may support many modes, however, they are often designed to support only the fundamental mode and our toolkit restricts to simulations of this case.
The nonlinearities of PICs are introduced with a small variation of the effective index in
\begin{equation*}
    \beta = \frac{2\pi \nu}{c_0}n^{eff}  \quad \rightarrow
    \beta = \frac{2\pi \nu}{c_0} \left( n^{eff} + n^{eff,NL} \right) \,.
\end{equation*}
In particular, when studying silicon-on-insulator PICs, the effects usually considered are the third-order nonlinear optical effects, the free carrier effects, and the thermo-optical effect, and so $n^{eff,NL}$ is a function of the field amplitude $\left|A\right|^2$ (optical intensity), the free carrier density $N$, and the temperature of the material $T$.
\begin{equation*}
    n^{eff,NL} = n^{eff,NL}( \, \left|A\right|^2, N, T) \,.
\end{equation*}

    The fields propagate symmetrically through the WGs following
    \begin{equation}
    	E_i^{in}(t) = E_j^{out}(t) e^{i\phi_k} \quad \text{and} \quad E_j^{in}(t) = E_i^{out}(t) e^{i\phi_k} \,,
    \end{equation}
    with
    \begin{equation}
    	\quad \phi_k = \frac{2\pi \nu}{c_0}\left( n^{eff}_k + n^{eff,NL}_k \right) L_k \,.
    	\label{eq:phase}
    \end{equation}
    $E(t)$ is the complex amplitude of the electromagnetic (EM) field, $\nu$ is its optical frequency, $c_0$ is the speed of light in vacuum, $L$ is the WG length, and $n^{eff}$ and $n^{eff,NL}$ are, respectively, the linear and nonlinear part of the effective index of the mode propagating within the WG.
    The user can define a new WG between existing ports using \texttt{add\_guide}.

\item The last important piece is the definition of one or more input ports to the whole system.
    In order to designate an input port, the user should call the function \texttt{add\_input} with the port index as an argument (see later).
    The input field of ports that are not connected with WGs or defined as inputs is assumed to be zero.
\end{enumerate}
When these three elements are correctly defined, PRECISE automatically creates the system of parametric equations representing the structure.\\
For example, by combining two $2\times2$ couplers and 2 WG segments, a microring (MR) resonator in the add-drop filter configuration is obtained, as depicted in \Cref{fig:ring-scheme}.
The set of equations derived from this optical system reads:
\begin{subequations}
	\begin{equation}
		E^{out}_1 = i k_1 E^{out}_8 e^{i\phi_1}
	\end{equation}
	\begin{equation}
		E^{out}_2 = \sqrt{1 - k_1^2} E^{in}_1 + i k_1 E^{out}_7 e^{i\phi_2}
	\end{equation}
	\begin{equation}
		E^{out}_3 = i k_1 E^{in}_1 + \sqrt{1 - k_1^2} E^{out}_7 e^{i\phi_2}
	\end{equation}
	\begin{equation}
		E^{out}_4 = \sqrt{1 - k_1^2} E^{out}_8 e^{i\phi_1}
	\end{equation}
	\begin{equation}
		E^{out}_5 = i k_2 E^{out}_4 e^{i\phi_2}
	\end{equation}
	\begin{equation}
		E^{out}_6 = i k_2 E^{out}_3 e^{i\phi_1}
	\end{equation}
	\begin{equation}
		E^{out}_7 = \sqrt{1 - k_2^2} E^{out}_3 e^{i\phi_1}
	\end{equation}
	\begin{equation}
		E^{out}_8 = \sqrt{1 - k_2^2} E^{out}_4 e^{i\phi_2}
	\end{equation}
\end{subequations}
where $E_j^{in/out}$ are the complex field amplitudes, $k_1$ and $k_2$ are the coupling constants, and the phases $\phi_1$ and $\phi_2$ are defined in \Cref{eq:phase}.

\begin{figure}[ht]
	\centering
	\includegraphics{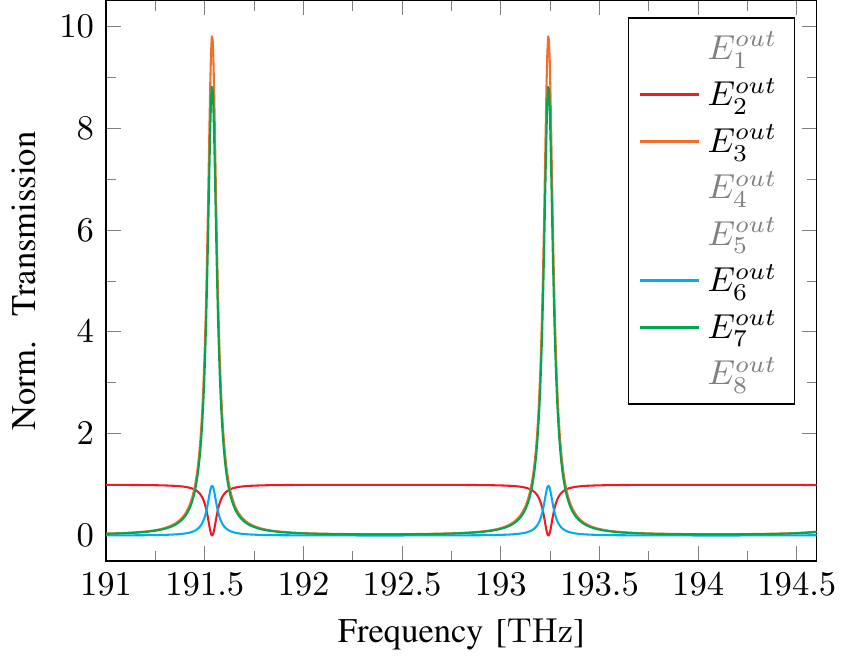}
	\caption{Microring resonator linear response.
		Transmission (power) associated with the field at the through port (red), and at the drop port (black), and (power) enhancement factors associated with the fields circulating in the right half (blue) and left half (yellow) ring respectively.
		The transmission correlated to the counter-propagating fields $E^{out}_1$, $E^{out}_4$, $E^{out}_5$, and $E^{out}_8$ are evaluated too, but are identically zero.
	}
	\label{fig:linear_ring}
\end{figure}

\vspace{1em}
Such description of electromagnetic fields is of general validity and allows both counter-propagating fields as well as refractive index variations due to nonlinear effects.\\
The direct solution of this symbolic system of equations is inefficient and becomes quickly intractable with the increasing number of PIC elements.
Thus, the system of symbolic equations is transformed in its matrix form and the numerical values of all parameters are substituted before its inversion.
To do this, PRECISE has a built-in function that takes as input the parameter's values and returns the numerical matrices describing the system.
These matrices are then fed to the linear system solver algorithm, which solves them by inversion and finds the complex amplitude of the fields at the output port of each and every connection node.\\
For example, in the case of the add-drop filter configuration with one input port (see \Cref{fig:ring-scheme}), the resulting function becomes:
\begin{multline}
	E^{out} = \\
	f \left( \nu, in_1, k_1, k_2, L_1, L_2,
	n^{eff}_1, n^{eff}_2, n^{eff,NL}_1, n^{eff,NL}_2 \right)
	\label{eq:pfe}
\end{multline}
and its output is
\begin{equation}
	E^{out} = \left[ E_{1}^{out}, \,\dots, \, E_{8}^{out} \right]^T,
\end{equation}
where $\nu$ is the optical frequency, $in_1$ is the complex field amplitude of the input signal, $k$ are the (two) coupling constants of the coupling regions, $L$ are the (two) lengths of each WG segment, the $n_{eff}$ and $n^{eff,NL}$ are the linear and nonlinear effective refractive indices of the (two) WG respectively, and $E_{1}^{out}, \dots, E_{8}^{out}$ are the complex amplitudes of EM fields at the output ports.
The resulting spectrum for this case, i.e. a single MR resonator in the add-drop filter configuration, is shown in \Cref{fig:linear_ring}.
Each line is obtained taking the square modulus of the given field and then is normalized for the square modulus of the input field, eventually obtaining the transmission: $||E_j||^2/||E_{in}||^2 \sim P_j/P_{in} \equiv T$.

The general procedure to create an optical system (all ports are indexed with positive integers) is:
\begin{lstlisting}[caption={[Algorithm] General optical system}, language=Matlab]
% initialize the class object
S = optical_system( );

% add nodes and connect them with links
% and waveguides; n is the number of ports.
S.add_node( n );
	
% add direct links between the ports, such as those
% found in a 2x2 coupler. It follows the equation
% ao=K*bi, with a (output) and b (input) port
% vectors and K a coefficient matrix.
S.add_link( a, b, K );

% connects the output port p1 to the input port p2
% and the output port p2 to the input port p1
% with a waveguide.
S.add_guide( p1, p2 );

% define the port p0 as inputs of the structure
S.add_input( p0 );

% generate the function that evaluates the fields
f = S.generate_pfe( );
\end{lstlisting}

The last command, \texttt{generate\_pfe}, generates the function that parametrically evaluates the output fields, as reported in \Cref{eq:pfe}.
This function is called \textit{parametric field evaluation} (PFE) and is generated automatically by the toolkit.
To simplify the modeling of PICs further, our toolkit enables a modular definition of new structures.
Hence, for example, a MR resonator structure in the add-drop filter configuration is easily defined with:
	
\begin{lstlisting}[caption={[Algorithm] MR definition}, language=Matlab]
S = optical_system( );
	
% add ring with 2 nodes, i.e. an add-drop filter
S.add_ring( 2 );

% designate port 1 as the input port
S.add_input( 1 );

f = S.generate_pfe( );
\end{lstlisting}
where the function \texttt{add\_ring} takes care of all the necessary steps, i.e. adding two nodes with four ports each and two WGs connecting the nodes together, eventually producing the structure represented in \Cref{fig:ring-scheme}.

\begin{sloppypar}
Examples of other functions are \texttt{add\_crow}, \texttt{add\_scissor}, \texttt{add\_taiji\_resonator}, and \texttt{add\_termination}, which add, respectively, a Coupled Resonators Optical Waveguide (CROW) structure (see \Cref{ssec:crow}), a Side Coupled Spaced Sequence of Optical Resonators (SCISSOR) structure \cite{mancinelli2011optical}, a Taiji MR (see \Cref{ssec:taiji}), and a reflective facet.
When possible, these functions define generic structures, where the number of sub-elements can be specified by the user setting an input argument to the function.
For example, in \texttt{add\_ring} the number of the MR coupling regions can be specified, while in \texttt{add\_crow} and \texttt{add\_scissor} the input parameter is the number of MRs that compose the structure.
\end{sloppypar}
	
\subsection{Temporal solution and nonlinear effects}
\label{ssec:temporal_solution_nonlinear}
In principle, the temporal evolution of the EM field in a PIC is the solution of a system of differential equations describing the relevant parameters: the optical field complex amplitude at each port, the temperature and the free carrier concentration in each WG, etc.
Such a method is computationally intensive as it requires taking time steps small enough to track the complex amplitude dynamic ($\sim \unit{ps}$).
On the other hand, if we assume that the optical field reaches the stationary state at each time step before any relevant change happens to either temperature or free carriers concentration, then we can consider an instantaneous transmission of the input optical field throughout the system, which is easily evaluated with the linear description obtained in the previous section.
To correctly describe the system in its steady-state, the $n^{eff,NL}$ is evaluated at each time step with the current values of temperature and free carriers.

The ODE solver algorithm used is MATLAB's own `\texttt{ode23}' algorithm, a non-stiff differential equations solver implementing an adaptive step size.
The different steps that the ODE solver carries out at each iteration are:
\begin{enumerate}
\item evaluation of refractive index variation $n_{0}^{NL}$ due to nonlinear effects involving temperature and free carriers \cite{chen2012bistability,mancinelli2013linear,borghi2016linear}:
    \begin{multline}
		n_{0}^{NL} = n_2\,I + i\frac{\beta_{TPA} c_0}{4\pi \nu} \,I + n_{TOE} \, \Delta T\\
		+\sigma_{FCD} \Delta N +i\frac{\sigma_{FCA} c_0}{4\pi \nu} \Delta N \,.
    	\label{eq:neff_NL}
    \end{multline}
    The effective refractive index change due to nonlinear effects is then obtained using the relation
    \begin{equation}
        n_{}^{eff,NL} = \Gamma \frac{n_g}{n_0} \, n_{0}^{NL},
    \end{equation}
    where $\Gamma$ is the optical mode confinement factor, defined as in \cite{borghi2016linear}.
\item evaluation of the optical fields using the parametric field evaluation function as seen in \Cref{eq:pfe}.
    Here the input signal at time $t$ becomes the input field amplitude $in$.
    \begin{equation}
        E^{out} = f \left( \nu, \{in\}, \{k\}, \{L\}, \{n^{eff}\}, \{n^{eff,NL}\} \right)
        \label{eq:pfe_nl_step}
    \end{equation}
with curly braces indicating a set of values.
\item evaluation of the differential change for the free carrier population and the temperature in each WG, adapted from \cite{johnson2006self}:
    \begin{subequations}
    		\begin{equation}
    			\frac{d}{dt}\Delta N = -\gamma_{FC} \, \Delta N
    			+ \left( \frac{c_0^2 \beta_{TPA} \Gamma_N}{2n_0 \hbar 2\pi \nu V_{eff}^2 } \right) \, U^2 \,,
    			\label{eq:deltaN}
    		\end{equation}
    		\begin{equation}
    			\frac{d}{dt}\Delta T = -\gamma_{TH} \, \Delta T + \frac{\Gamma}{MC_P} \, P_{abs}\left(U\right) \,.
    			\label{eq:deltaT}
    		\end{equation}
    		\label{eq:system}
    \end{subequations}
\end{enumerate}
$I=\frac{1}{2}c_0\epsilon_0E^2$ is the irradiance of the field circulating in the WG, where $\epsilon_0$ is the vacuum permittivity, and $U = Ln_g E^2/ c_0$ its energy, evaluated from the fields $E$, while $\Delta T$ and $\Delta N$ are the deviations of the temperature and the Free Carrier (FC) density from their equilibrium value, respectively.
Moreover, the power generating heat is (adapted from \cite{johnson2006self}):
\begin{equation}
	P_{abs} = 2 \left( \frac{\sigma_{FCA} c_0 \Gamma_N}{2n_0} \Delta N
	+ \frac{ \beta_{TPA} c_0^2 }{ n_0^2 \,V_{eff} } \,U
	+ \gamma \right) \,U \,.
	\label{eq:power}
\end{equation}
The other parameters are described in \Cref{tab:parameters}.

\captionsetup[table]{font+=sc, labelsep=newline, justification=centering}
\begin{table*}[ht]
    { \centering
	\caption{
	    List of the parameters that appear in \Cref{eq:system} with their description and definition.
    }
	\label{tab:parameters}
\sisetup{table-format = 2.4, table-alignment-mode = format}
\begin{tabular}{@{}
                c
                c
                S[table-format=+1.3e+3]
                >{\collectcell\unit}l<{\endcollectcell}
                c
                |
                c
                c
                S[table-format=+1.3]
                >{\collectcell\unit}l<{\endcollectcell}
                c
                @{}
}
    \toprule
	{Quantity} & {Description} & {Value} & {Unit} & {Source} & {Quantity} & {Description} & {Value} & {Unit} & {Source}\\
	\midrule
	$n_0$ & silicon refractive index & 3.485 & & \cite{johnson2006self} &
	    $\gamma$ & linear absorption & 1.7 & \GHz & \cite{johnson2006self}* \\
	$n_g$ & group index & 4 & & FEM &
	    $\gamma_{FC}$ & free carrier decay rate & 0.3 & \GHz &  \cite{mancinelli2013linear}* \\
	$n_2$ & Kerr effect & 4.5e-9 & \square\um\per\mW & \cite{chen2012bistability} &
	    $\gamma_{TH}$ & thermal decay rate & 0.045 & \GHz &  \cite{mancinelli2013linear}* \\
	$\beta_{TPA}$ & two photon absorption & 8.1e-9 & \um\per\mW
	& \cite{johnson2006self,chen2012bistability}* &
	    $\Gamma$ & field conf. factor & 0.99 & & FEM \\
	$n_{TOE}$ & thermo-optic effect & 1.86e-4 & \per\K & \cite{johnson2006self} &
	    $\Gamma_N$ & free carrier conf. factor & 0.99 & & FEM \\
	$\sigma_{FCD}$ & free carrier dispersion & -4.0e-9 & \um\cubed
	& \cite{johnson2006self}* &
	    $V$ & volume & 2.474 & \um\cubed & geom. \\
	$\sigma_{FCA}$ & free carrier absorption & +1.4e-9 & \square\um
	& \cite{johnson2006self,chen2012bistability}* &
	    $V_{eff}$ & effective volume & 3.093 & \um\cubed & FEM \\
	$\alpha$ & linear losses & 2e-4 & \dB\per\um  & \cite{vlasov2004losses}* &
	    $M$ & mass & 5.762 & \pg & geom.\\
	$C_P$ & silicon specific heat & 0.710 & \J\per\g\per\K & \cite{johnson2006self}\\
	\bottomrule \addlinespace[\belowrulesep]
\end{tabular} }
	{ \footnotesize
	    The values whose sources are marked with an asterisk * should be considered of the same order of magnitude of the given sources.
	    The left half of the table contains values typical of silicon photonics structure, whereas the right half contains quantities whose values have been adapted to describe the structure studied in \Cref{ssec:nonlinear_ring}.
    }
\end{table*}
	
Notice that if the first two terms in \Cref{eq:neff_NL} are negligible, points 1 and 2 are evaluated only once.
Otherwise, their evaluation should continue until self the consistency condition is found.
In this case, however, the computational complexity becomes similar to that of the CMT and so does the simulation time.\\
In essence, the dynamic of effects faster than roughly ten times the cavity lifetime (see the more in-depth discussion in the \Cref{sec:benchmarks}) cannot be accurately described by PRECISE (i.e. as the Kerr effect and the Two Photon Absorption (TPA) contributions) since they are neglected during the updates of the optical fields in \Cref{eq:neff_NL}.
Yet, for silicon photonics, these effects are often orders of magnitude smaller than free carriers and thermal dispersion ones.
However, TPA is still considered in \Cref{eq:power} for the evaluation of the total absorbed power.

\subsection{Parametric solutions}
\label{ssec:parametric_solutions}
An important goal of this toolkit is to allow the analysis of systems by exploring their parameter space, which may have several dimensions even for small systems (i.e. for a single MR we might consider: MR radius's, wavelength, losses, coupling strengths, etc.).
For this reason, in addition to the fact that the functions evaluating the linear field propagation and its nonlinear evolution are parametric, the toolkit helps the user to explore the parameter space methodically.
A specific class (\texttt{dataloader}) stores all the values that any parameter assumes, such that any desired combination can be retrieved via the indexing of a single integer number, going from $1$ to the total number of combinations $N_C$.
This set of parameters is then given as arguments to the function representing the system that, in turn, is evaluated by the ODE solver.

The benefits of this technique are twofold.
The first advantage is that it is straightforward to run parallel evaluations of many configurations, using Matlab's \texttt{parfor} loop, thus increasing the overall computational efficiency.
The second one is that each temporal solution is analyzed right after its evaluation and, once the interesting figures-of-merit are extracted, the data of those time series can be eliminated.
A given solution can be re-generated at any time by simply looking at its scalar index.
\Cref{ssec:data_analysis_and_visualization} will explain this more in detail.
These three characteristics, together, contribute to a reduced memory footprint and a faster overall execution time, particularly for large parameter spaces.

\subsection{Data analysis and visualization}
\label{ssec:data_analysis_and_visualization}
For each combination of parameters, the time evolution of several quantities is evaluated.
PRECISE calculates the complex field amplitude at each port and the deviation of the temperature and the FC density from their equilibrium values.
For example, in the case of a MR, the toolkit evaluates the evolution of twelve time series: 8 fields, 2 temperatures, and 2 free carriers, since each MR is described by two WG segments.
The output dataset would then be three-dimensional, with size $N_C \times N_S \times N_t$, where $N_C$ is the number of combinations, $N_S$ is the number of time series per combination, and $N_t$ is the number of samples in each time series.
When long traces are evaluated or when the $N_C$ is large, the amount of data that is generated becomes quickly intractable.

Since in almost any case $N_t > N_C > N_S$, it is preferable to synthesize the results by extracting a (scalar) figure-of-merit from each time series.
The resulting dataset is bi-dimensional and has size $N_C \times N_S$.
Apart from the most straightforward (i.e. minimum, maximum, and average values, from which the signal visibility is derived), custom functions are defined to evaluate more complex figures of merit, such as the period of the most significant FFT component (\texttt{periodFFT}), the number of the extremal points (\texttt{extremals}), and the unique values of the extremal points (non-scalar, \texttt{extremals\_counts}).
	
Even though the figure of merit condenses the temporal dimension into a scalar value, the dimension along which the results of different combinations are stacked represents the parameter space explored by all the combinations.
This is usually n-dimensional and hence requires \textit{ad-hoc} visualizations.
Our toolkit provides several functions and classes to properly visualize this information.
A first class (\texttt{dashboard}) offers a way to observe the temperature and the FC density average, maximum, and minimum values for all the combinations evaluated.
It also plots their time evolution as well as the absolute value of the complex field amplitude for a specific combination of parameters selected by the user.
A separate class (\texttt{array\_plot}) adds the possibility to observe any generic figure of merit for all the combinations related to two parameters at a time, e.g., the frequency against the input power or the coupling coefficient against the MR’s radius.
All ‘views’ can be updated with different parameters and/or for different combinations.
	
\section{Use cases}
In this section, we present several use cases to demonstrate the capabilities of our toolkit.
The first two examples rely on linear spectra calculation of devices with peculiar structures.
The third case describes the temporal dynamics of a nonlinear resonator.

\subsection{Linear spectrum}
The simplest way of using PRECISE is to simulate linear transmission spectra of integrated optical structures.
The outcome of this simulation is a set of complex field amplitudes, from which the transmission and enhancement factor values are easily extracted.
We propose two examples of resonant devices with peculiar structures and compare the modeling results with experimental data.

\begin{figure}[ht]
    \centering
    \includegraphics{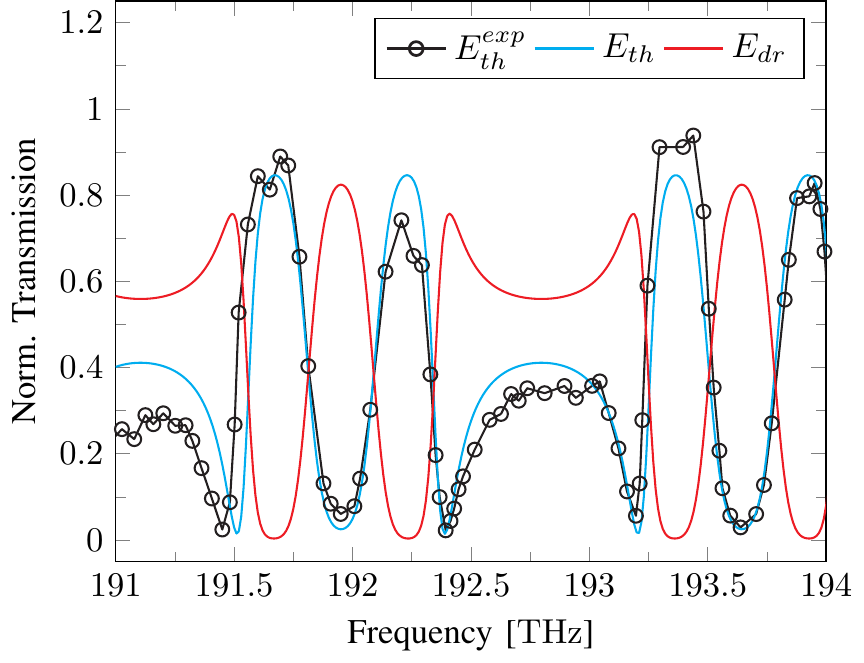}
    \caption{ \label{fig:linear_crow}
        CROW structure (interferometric band interleaver) linear response.
        Transmission (power) of the fields at the through port (cyan) and at the drop port (red).
        The experimental values (black circles) of the transmission in the drop port are shown as well.
    }
\end{figure}

\subsubsection{Coupled Resonators Optical Waveguide structure}
\label{ssec:crow}
A CROW is composed of side-coupled multiple MRs, sandwiched between two bus WGs in ``add-drop'' configuration \cite{poon2006transmission}.
We show the result of a simulation that considers a CROW made of four rings with the input signal entering from both ``input'' and ``add'' ports with a $\pi/2$ phase difference.
This particular configuration is known as ``interferometric band interleaver'' \cite{melloni2002synthesis,mancinelli2013interferometric} and its scheme is reported in \Cref{fig:crow-scheme}.
In \Cref{fig:linear_crow} one can observe that the model correctly predicts the free spectral range of the resonances and their complex shape due to the interaction between the rings.
We coupled the PRECISE solver with an optimization routine base on Matlab's `\texttt{particleswarm}' to find the optimal set of parameters describing the CROW spectral response, considering all radii and all coupling coefficients equal.
The outcome of the optimization correctly describes the free spectral range of the resonance as well as their complex shape due to the interaction between the rings, as reported in \cite{mancinelli2013interferometric,mancinelli2011optical}.
The parameters are reported in \Cref{tab:crow-parameters}.

\begin{figure}[ht]
    \centering
    \includegraphics{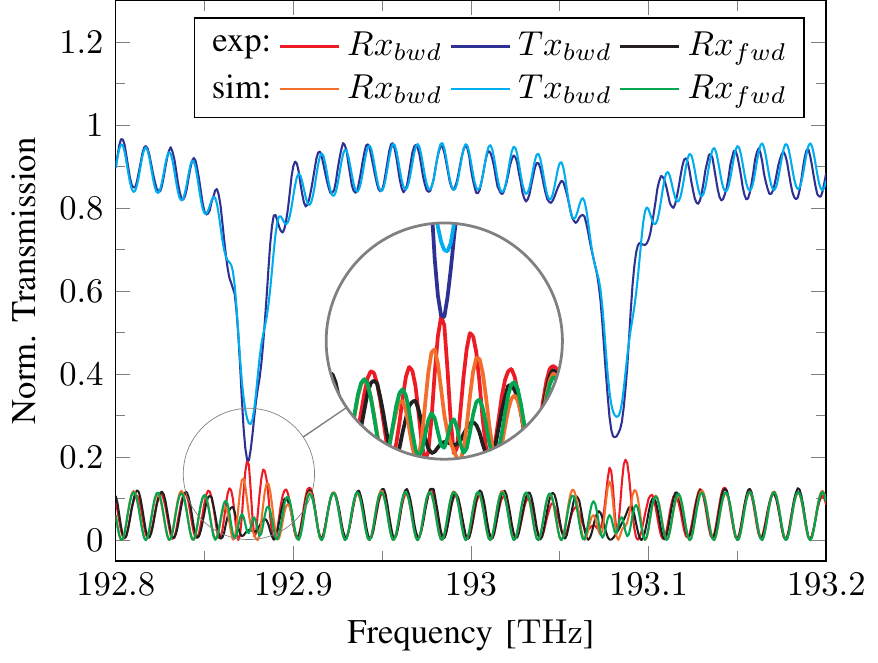}
    \caption{ \label{fig:linear_taiji}
        Taiji MR resonator linear response.
        Experimental data for backward direction ($bwd$), reflection $Rx$ (red) and transmission $Tx$ (blue), and forward direction ($fwd$) reflection $Rx$ (black).
        Numerical data for backward direction ($bwd$), reflection $Rx$ (yellow) and transmission $Tx$ (brown), and forward direction ($fwd$) reflection $Rx$ (green).
    }
\end{figure}

\subsubsection{Taiji microring resonator structure}
\label{ssec:taiji}
This example describes a Taiji MR resonator composed of bus WGs and reflective facets.
A complete in-deep description of this device can be found in \cite{calabrese2020unidirectional}.
Briefly, the Taiji is a particular MR configuration described by a node with two ports designed such that the reflection of either port depends on which port the input field enters the device itself (\Cref{fig:taiji-scheme}).
The device is reciprocal as the total transmission of the structure is the same for both propagation directions, but it shows an asymmetric reflection response.
Our model correctly describes all the Taiji spectral features: the asymmetric reflection, the symmetric transmission, as well as the Fabry-Perot effect of the facets.
\Cref{fig:linear_taiji} shows the back-reflection (Rx) and the transmission (Tx) for the backward ($bwd$) mode and the transmission only (Tx) for the forward mode ($fwd$), for both the numerical model and the experimental data.
In the Taiji the total reflection is due to counter-propagating fields, thus the good agreement with the experimental data confirms that those fields are correctly handled by the toolkit.
Similar to the CROW case, we employed an optimization routine to find the set of parameters that best approximate the spectral response.
\Cref{tab:taiji-parameters} reports the simulation parameters

\begin{figure*}[ht]
    \centering
    \includegraphics{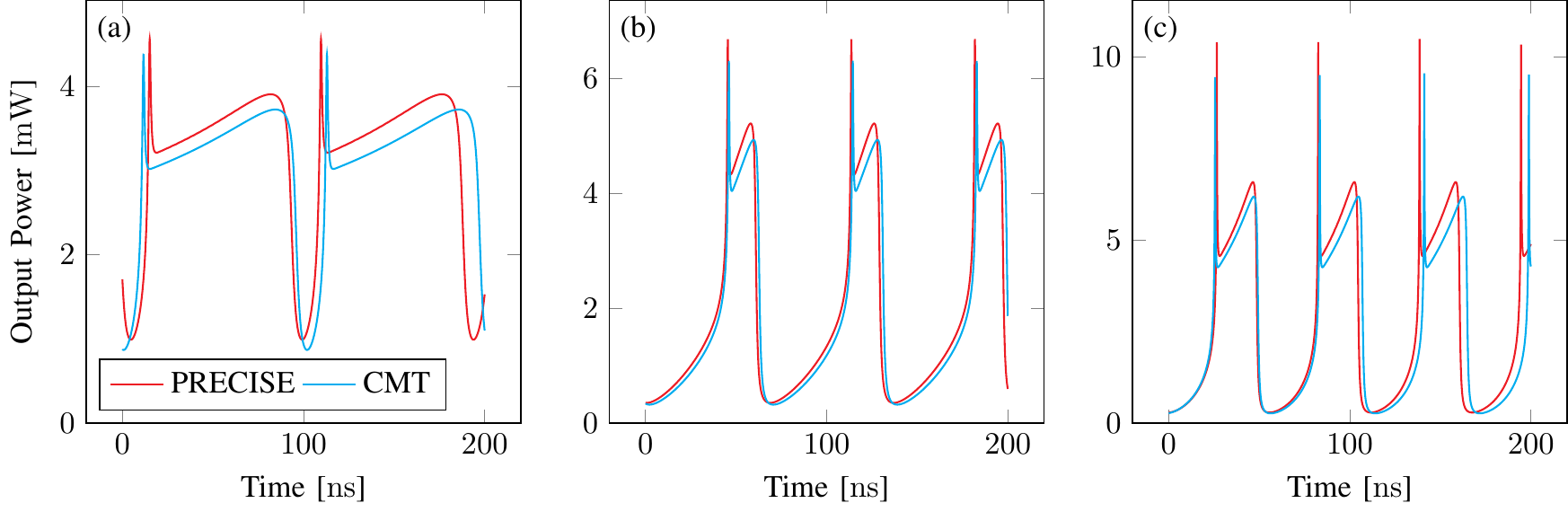}
    {\phantomsubcaption\label{fig:tracea}}
    {\phantomsubcaption\label{fig:traceb}}
    {\phantomsubcaption\label{fig:tracec}}
    \caption{
        \label{fig:nonlinear_comparison}
        Microring nonlinear response in the drop port ($||E_6^{out}||^2$) for three different combinations of input power and frequency.
        The curves shown above are obtained from longer simulations and the time interval is chosen so that the response can be considered as a stationary state.
    }
\end{figure*}
\begin{figure*}[ht]
	\centering
	\includegraphics{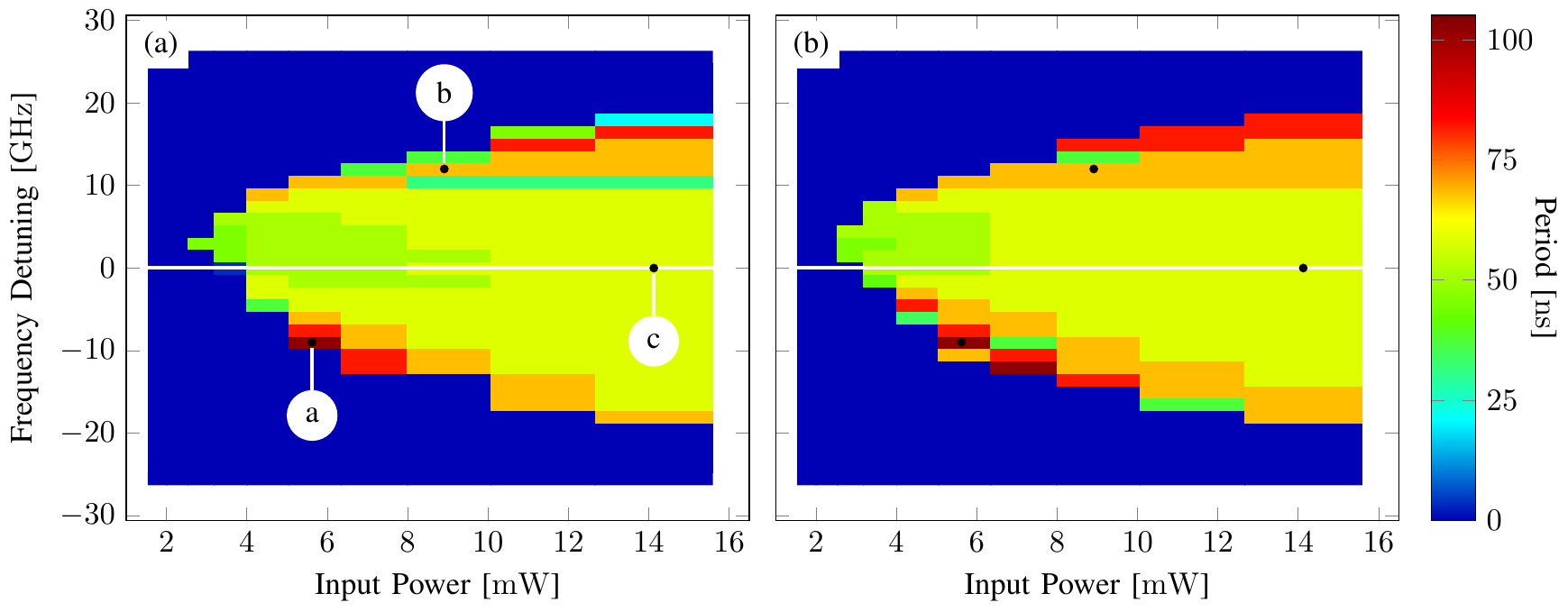}
    {\phantomsubcaption\label{fig:periods_PRECISE}}
    {\phantomsubcaption\label{fig:periods_CMT}}
	\caption{
	    \label{fig:periods}
	    Instability region colored with the period of the most important Fourier-component observed on the drop port ($\left\| E_{6}^{out} \right\|^2$).
	    The region where the microring does not self-pulse is shown as having a \qty{0}{\ps} period.
	    The horizontal white lines highlight the microring resonance frequency, at $0 GHz$ detuning.
	    \subref{fig:periods_PRECISE} Shows the values for PRECISE, whereas \subref{fig:periods_CMT} shows the results for the corresponding CMT implementation.
	    The highlighted points refer to the curves shown in \Cref{fig:nonlinear_comparison}.
    }
\end{figure*}

\subsection{Temporal dynamic of nonlinear microring resonator}
\label{ssec:nonlinear_ring}
Considering a single MR, for high enough fields intensity, we expect to observe a region around the MR's resonance exhibiting the so-called ``self pulsing'' behavior \cite{van2012simplified}.
The ``self pulsing'' arises as a dynamical modulation of the optical field at the output of the MR as a response to a continuous wave input signal.
The specific shape and period of the output signal depend non trivially on several parameters, e.g., device materials and geometry, injected optical power, or detuning from the resonance.
For the case reported in \Cref{fig:nonlinear_comparison}, we consider a single ring resonator, whose radius is \qty{7}{\um} and whose coupling coefficients are $k_1 = k_2 = 0.17$, probed with a signal of constant optical power.
The values of all the other parameters are the same reported in \Cref{tab:parameters}.
The red line of \Cref{fig:nonlinear_comparison} shows the NL response of the ``drop'' port ($||E_6^{out}||^2$) for three different combinations of input power and frequency detuning.
In the same graphs, the temporal evolution calculated with a CMT model for the same input parameters is shown with the blue line.
Our code correctly describes both the frequency and the amplitudes of the signal modulation, with only small differences compared to the CMT model.
These are due to the different approximations considered by CMT and by our code.
One of the most important is that while the resonance frequency emerges automatically in our code, it is imposed as a parameter in the CMT model, thus resulting in a possible misalignment of the frequencies evaluated in PRECISE and CMT.
Another important difference is that in PRECISE the microring cavity is composed of separated waveguide segments connecting at the coupling regions, while in the CMT it is a unique waveguide.
This enables PRECISE to have different values for the physical quantities in each section of the microring.
Moreover, the dispersion of $n^{eff}$ and the waveguide losses are approximated differently in the two codes.
Overall the matching of the two codes is good and it is also confirmed by the results described in the next paragraph.

\begin{figure*}[ht]
	\centering
	\includegraphics{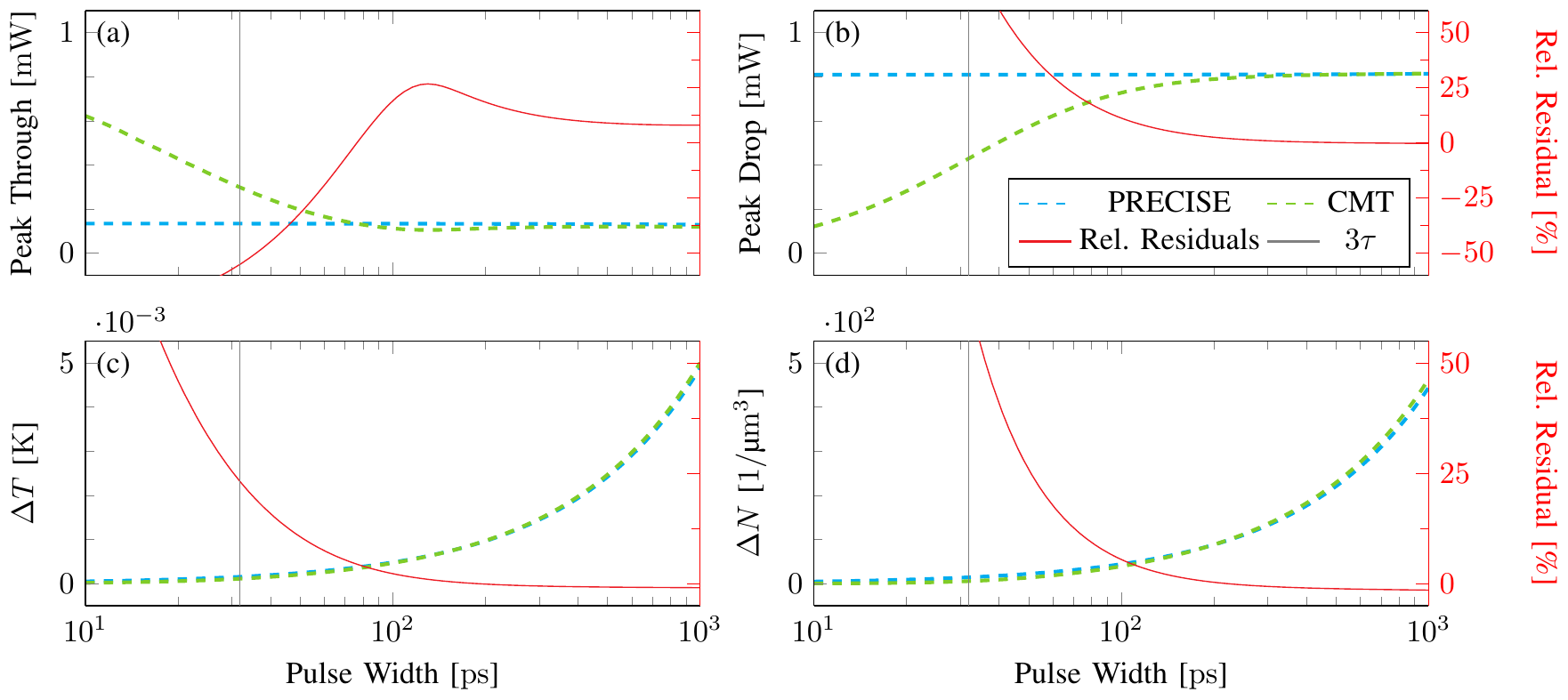}
	{\phantomsubcaption\label{fig:peak_through}}
	{\phantomsubcaption\label{fig:peak_drop}}
	{\phantomsubcaption\label{fig:mean_dt}}
	{\phantomsubcaption\label{fig:mean_dn}}
	\caption{
    	Comparison between PRECISE and CMT models for the propagation of Gaussian pulses of varying width through an add-drop filter near its resonance frequency.
    	\subref{fig:mean_dt} MR temperature difference, \subref{fig:mean_dn} Free carrier concentration, \subref{fig:peak_through} peak power in the \textit{Through} port, and \subref{fig:peak_drop} peak power in the \textit{Drop} port.
    	Vertical black lines point to three times ($3\tau$) the cavity lifetime.
    	Continuous blue lines indicate the results of PRECISE, while dashed green lines are the CMT model (left axis).
    	 The red, solid line reports the residuals (right axis).
    }
	\label{fig:benchmarks1}
\end{figure*}

\Cref{fig:periods} shows the stability maps obtained analyzing the complex field amplitude of the ``drop'' port ($\left\| E_{6}^{out} \right\|^2$) in a MR resonator as a function of the input power and of the frequency detuning.
The maps report the principal non-zero Fourier component of the spectra of the MR in the self-pulsing regime: panel \subref{fig:periods_PRECISE} refers to our code, while panel \subref{fig:periods_CMT} reports the results for the CMT one.
The three points highlighted in the map refer to the spectra reported in \Cref{fig:nonlinear_comparison}.

As expected, the system shows self-pulsing behavior in a limited space region around the resonance frequency and at input power above a certain threshold.
The actual shape of this region is heavily dependent on the details of the system, as it is defined by the interplay between the optical energy in the cavity, the free carriers population, and the temperature of the material.
The shape of the self-pulsing region is very similar between the two implementations, with only minor differences in the actual period of the oscillations.

\section{Benchmarks}
\label{sec:benchmarks}
We perform two different benchmarks to test both the accuracy and the speed of the library.

The accuracy of the model has been tested by comparing the propagation of optical pulses of different widths sent through a single MR in add-drop configuration near its resonance frequency (see \Cref{fig:benchmarks1}).
The four panels describe the variation of some of the fundamental system parameters: (top-left) the peak power in the T-port, (top-right) peak power in the D-port, (bottom left) MR temperature increase, and (bottom-right) free carrier concentration increase.
For all simulations the maximum peak intensity has been kept constant, thus the overall total pulse power increases proportionally to its duration.
The main difference between PRECISE and the CMT is visible in the dynamics of the MR loading, where the CMT correctly describes the dynamics of the energy pile up in the cavity, while our model assumes a steady-state condition.
On the other hand, both $\Delta T$ and $\Delta n$ are correctly described.
The time scale that PRECISE correctly handles can be inferred by using the cavity lifetime as a scaling factor.
The vertical black lines indicate the cavity lifetime and it is clear that PRECISE accurately describes the system dynamics for timescale one order of magnitude greater than the cavity lifetime.
For dynamics with characteristic times slower than $\tau$, the difference between the two models is around $2\%$.

\begin{figure*}[ht]
	\centering
	\includegraphics{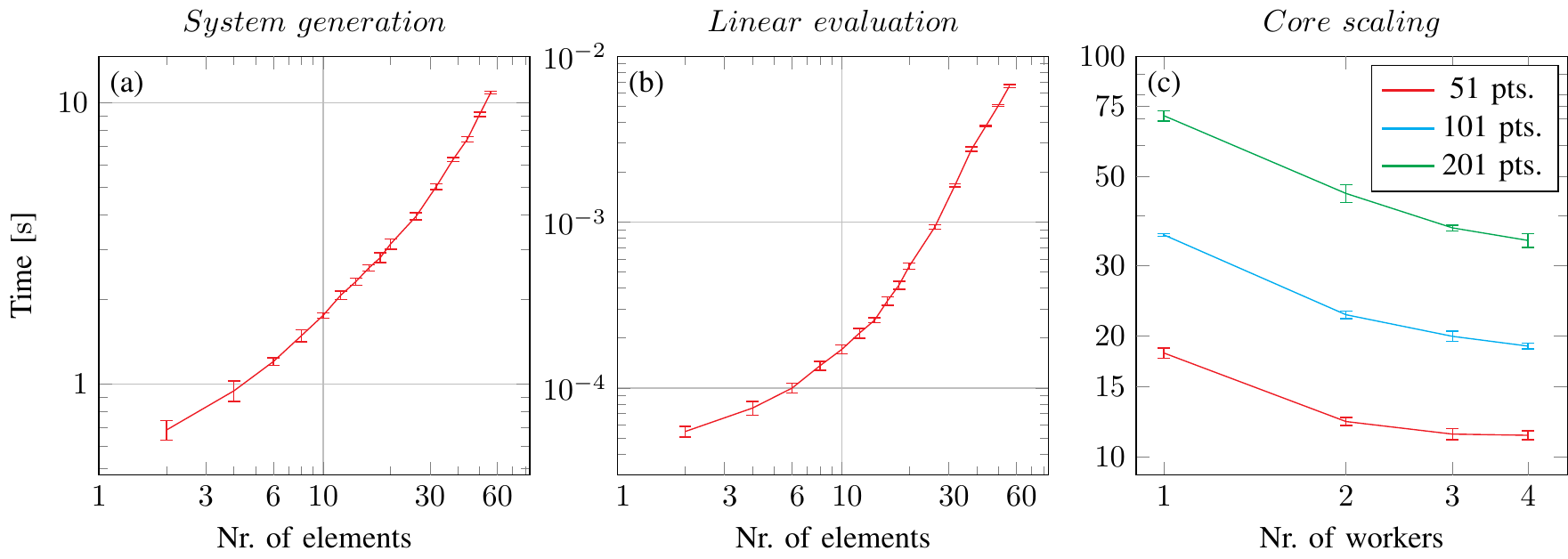}
    {\phantomsubcaption\label{fig:sys_generation}}
    {\phantomsubcaption\label{fig:linear_eval}}
    {\phantomsubcaption\label{fig:core_scaling}}
	\caption{
    	Performance analysis for the CROW structure, carried out by taking the average and the standard deviation on ten consecutive runs.
    	Each additional ring in the CROW adds two elements to the system.
    	\subref{fig:sys_generation}
    	    Time necessary to initialize the system and generate the corresponding symbolic equations.
    	\subref{fig:linear_eval}
    	    Time necessary for a single linear evaluation of the EM fields.
    	    Obtained from a batch evaluation of 1001 points.
    	\subref{fig:core_scaling}
    	    Time necessary to simulate the nonlinear evolution of a CROW with two rings for 51, 101, and 201 frequency points as a function of the number of parallel workers (processes).
            Matlab's own \texttt{parpool} initialization time is neglected.
            Time may vary with the task.
    }
	\label{fig:benchmarks}
\end{figure*}

Finally, we run a benchmark to show the performance of PRECISE by running a suite of simulations on CROW structures of increasing complexity.
We choose the CROW as it is composed of multiple rings and so by a number of elements (WGs and ports) proportional to the number of the rings.
All the simulations have been run with Matlab 2020a on a laptop PC equipped with an AMD Ryzen 5 2500U (15W, 4 cores, 8 threads) and 8 GB RAM.
Performance metrics are shown in \Cref{fig:sys_generation}, where all graphs are plotted in a log-log scale and the values are obtained with a statistic of 10 consecutive runs.
\Cref{fig:sys_generation} shows the time required to initialize the symbolic system and to obtain the parametric function that evaluates the EM fields at each port as a function of the number of elements in the system.
The system generation time shows a polynomial trend which sees the increase of time needed by one order of magnitude for an increase in the number of elements by almost two orders of magnitude.
\Cref{fig:linear_eval} shows the time required for a single evaluation of the parametric function, where the actual value has been obtained from a batch evaluation of 1001 points.
In this case, the linear evaluation time shows initially a polynomial trend which then seems to approach a power-law trend for higher number of system elements.
Finally, \Cref{fig:core_scaling} shows the scaling of the evaluation time on the number of processes, with three lines representing the parallel evaluation of the system in 51, 101, and 201 frequency points, respectively.
With just four workers we see a reduction of the computation time of $38\%$ for 51 points, $47\%$ for 101 points, and $51\%$ for 201 points.
In a real case scenario, the user would probably simulate several hundreds or thousands of combinations.
In this case, we can expect the performance improvement obtained with the parallel evaluation to be much more beneficial than in the small benchmarks presented here.

\section{Conclusion} 
PRECISE is a highly modular toolkit that allows simulating time-resolved linear and nonlinear regimes of large PICs.
The software is based on a set of general differential equations that describe any passive PICs with the only approximation of steady-state condition of the electric field within each block at each time step.
The engine is highly expandable to enable the description of different systems of equations (i.e. in \cite{Borghi21} a double decay time has been introduced to correctly describe the free carriers' decay dynamics).
Our results demonstrate that the steady-state approximation is rather general and enables the modeling of large circuits without sacrificing the accuracy of the simulations.
Furthermore, the toolkit allows for efficient parallelization of workloads and enables to sweep over broad parameter spaces.
The validity of the library is demonstrated by the comparison with another analytical model (CMT) and, more importantly, with experimental results with peculiar PIC devices such as the CROW or the Taiji MR.
The dynamic of nonlinear effects is also demonstrated with the analysis of the self-pulsing regime.

\appendix[Fit parameters of the linear spectra]

The following tables contain the fit parameters for the models described in \Cref{ssec:crow} and \Cref{ssec:taiji}.
The data can be easily reproduced by feeding the set of parameters to the `\texttt{linear\_crow}' and `\texttt{linear\_taiji}' functions found in `\texttt{precise.examples}'.

\begin{table}[ht]
    { \centering
    \caption{CROW simulation parameters
    }
    \label{tab:crow-parameters}
    \begin{tabular}{llll}
	    \toprule
		{Fitting Parameters} & {Values} & {Fixed Parameters} & {Values}\\
		\midrule
		$n_{eff}$ & 2.6391 & loss & \qty{2}{\dB\per\cm} \\
		$n_g$ & 4.3416 & ring number, $n$ & 4 \\
		$k$ & 0.83645 & & \\
 		radius, $R$ & \qty{3.2544}{\um} & WG segments, $L$ & $2\pi R+\qty{10}{\um}$ \\
		\bottomrule \addlinespace[\belowrulesep]
    \end{tabular} }
	{ \footnotesize
        $k$ are field coupling coefficients and their indexes identify the ports of the node in which they are used.
        $L$ are the WG lengths and the indexes identify the port at the ends of the WG.
        All indexes refer to \Cref{fig:crow-scheme}.
    }
\end{table}
\begin{table}[ht]
    { \centering
    \caption{Taiji MR simulation parameters
    }
    \label{tab:taiji-parameters}
    \begin{tabular}{llll}
	    \toprule
		{Fitting Parameters} & {Values} & {Fixed Parameters} & {Values}\\
		\midrule
		$n_{eff}$ & 1.6623 & linear loss & \qty{2.16}{\dB\per\cm} \\
		$n_g$ & 1.8224 & $L_{1-13}$ & \qty{3.3034}{\mm} \\
		$k_{1,2,3,4}$ & 0.29751 & $L_{2-15}$ & \qty{2.6715}{\mm} \\
		$k_{5,6,7,8} = k_{9,10,11,12}$ & 0.38599 & $L_{3-8}=L_{4-11}$ & \qty{0.202}{\mm} \\
		$k_{F1}$ & 0.20305 & $L_{7-12}$ & \qty{0.391}{\mm} \\
		$k_{F1}$ & 0.20328 & $L_{6-10}$ & \qty{0.611}{\mm} \\
		         &         & $L_{5-17}=L_{9-19}$ & \qty{0.118}{\mm} \\
        \bottomrule \addlinespace[\belowrulesep]
    \end{tabular} }
    { \footnotesize
        $k$ are field coupling coefficients and their indexes identify the ports of the node in which they are used.
        $L$ are the WG lengths and the indexes identify the port at the ends of the WG.
        All indexes refer to \Cref{fig:taiji-scheme}.
    }
\end{table}

\section*{Code availability}
The PRECISE toolkit for Matlab is freely available at \url{https://gitlab.com/erc-backup/precise} under the MIT license.
Its documentation can be found at \url{https://gitlab.com/erc-backup/precise/-/wikis/home}.
All contributions to the toolkit and its documentation are welcome.

\section*{Acknowledgments}
The authors would like to thank Stefano Biasi and Riccardo Franchi for their support and the fruitful discussions on the matter of the Taiji resonator.


\section*{Declaration of interests}
The authors declare that they have no known competing financial interests or personal relationships that could have appeared to influence the work reported in this paper.


\bibliographystyle{IEEEtran}
\bibliography{main}

%
%
%
%
%
%
%

\end{document}